\documentclass[prb,twocolumn,showpacs,superscriptaddress]{revtex4-1}

\usepackage[latin1]{inputenc}
\usepackage[french]{babel}
\usepackage{graphicx}
\usepackage{amssymb}
\usepackage{amsmath}
\usepackage{ulem}

\usepackage{color}

\begin{document}

\def \aI3 {$\mathrm{\alpha-(BEDT-TTF)_{2}I_3}$\,}
\def \SiO2 {$\mathrm{SiO_2}$\,}
 
\def\tc{$T_{c}$\,}
\def\pc{$P_{c}$\,}

\title{Evidence for the coexistence of Dirac and massive carriers in  \aI3 under hydrostatic pressure}
\author {M.Monteverde}
\author {M.O. Goerbig}
\author {P. Auban-Senzier}
\author {F.Navarin}
\author {H.Henck} 
\author {C.R. Pasquier}
\affiliation{Laboratoire de Physique des Solides, UMR 8502-CNRS, Univ.Paris-Sud, Orsay, F-91405, France}
\author{C.M\'ezi\`ere}
\author{P.Batail}
\affiliation{MOLTECH-Anjou, UMR 6200, CNRS-Universit\'e d'Angers, B\^at. K, Angers, F-49045, France}


\begin{abstract}

Transport measurements were performed on the organic layered compound \aI3 under hydrostatic pressure. The carrier types, densities and mobilities are determined from the magneto-conductance of \aI3 . While evidence of high-mobility massless Dirac carriers has already been given, we report here, their coexistence with low-mobility massive  holes. This coexistence seems robust as it has been found up to the highest studied pressure. Our results are in agreement with recent DFT calculations of the band structure of this system under hydrostatic pressure. A comparison with graphene Dirac carriers has also been done.
\end{abstract}

\date{\today}
\maketitle

\section{Introduction}

The layered organic material \aI3 (aI3), which has been studied since the 1980s,\cite{Mishima95,Tajima00,Tajima06,Tajima09,Kino06}
has recently attracted renewed interest because it reveals low-energy massless Dirac fermions under hyrdostatic 
pressure ($P>1.5$ GPa).\cite{Kobayashi07} Compared to graphene, certainly the most popular material with low-energy Dirac fermions\cite{Novoselov05}
or electronic states at the surface of three-dimensional topological insulators,\cite{TI} aI3 is strikingly different in several respects. 
Apart from the tilt of the Dirac cones and the anisotropy in the Fermi surface,\cite{Kobayashi07,Goerbig08} its average Fermi velocity is roughly
one order of magnitude smaller than that in graphene. This, together with an experimentally identified low-temperature charge-ordered phase at
ambient pressure,\cite{Mishima95,Tajima00} indicates the relevance of electronic correlations. Indeed, because the effective coupling constant 
for Coulomb-type electron-electron interactions is inversely proportional to the Fermi velocity, it is expected to be ten times larger in aI3 than in
graphene. The material aI3 thus opens the exciting prospective to study strongly-correlated Dirac fermions that are beyond the scope of graphene electrons.\cite{CastroNeto09}. 

Another specificity of aI3 is the presence of additional massive carriers in the vicinity of the Fermi level, as 
recently pointed out in ab-initio band-structure calculations.\cite{Alemany12} However, the interplay between massless Dirac fermions and 
massive carriers has, to the best of our knowledge, not yet been proven experimentally. Finally, one should mention a topological merging
of Dirac points that is expected for high but experimentally accessible pressure.\cite{Kobayashi07,Montambaux09}

Here, we present magneto-transport measurements of aI3 crystals under hydrostatic pressure larger than $1.5$ GPa where Dirac carriers are present. 
We show not only the existence of high-mobility Dirac carriers as reported elsewhere,\cite{Tajima06,Tajima11,Tajima12} but we
prove also experimentally the presence of low-mobility massive holes, in agreement with recent band-structure calculations.\cite{Alemany12}
The interplay between both carrier types at low energy is the main result of our studies.
Furthermore, we show that the measured mobilities for the two carrier types hint at scattering mechanisms due to strongly screened 
interaction potentials or other short-range scatterers.

The remainder of the paper is organized as follows. In Sec. \ref{sec:1}, we present the experimental
set-up and the results of the magneto-transport measurements (Sec. \ref{sec:1.1}) under hydrostatic pressure. The subsection \ref{sec:1.2} is devoted to a discussion of the temperature dependence of
the carrier densities, in comparison with the model of (A) massless Dirac fermions and (B) massive 
carriers. Furthermore thermopower measurements are presented to corroborate the two-carrier scenario.
The measured temperature dependence of the extracted carrier mobilities is exposed in Sec. 
\ref{sec:1.3}, and a theoretical discussion of the experimental results, in terms of short-range (such 
as screened Coulomb) scatterers may be found in Sec. \ref{sec:2}. We present our conclusions and
future perspectives in Sec. \ref{sec:3}.

\section{Experimental evidence for coexisting Dirac and massive carriers}
\label{sec:1}

The single crystals of aI3 used in our study have been synthesized by electro-crystallization. Their typical size is $1$ mm$^2$ ($ab$ plane) x $20\:\mu$m ($c$ direction). Six $100$ nm thick gold contacts were deposited by Joule evaporation on both sides of the sample, allowing for simultaneous longitudinal and transverse resistivity measurements. A picture of one of the three samples studied is shown in the inset of figure \ref{magneto}. The resistivities were measured using a low-frequency ac lock-in technique. The magnetic field $H$, oriented along the $c$ axis, was swept between $-14$ and $14$ T at constant temperature between $50$ and $1.5$ K. To account for alignment mismatch of patterned contacts, the longitudinal (transverse) resistivity has been symmetrized (anti-symmetrized) with respect to the orientation of $H$ to obtain even [$\rho_{xx}(H)$] and odd [$\rho_{xy}(H)$] functions respectively. Hydrostatic pressure was applied at room temperature in a NiCrAl clamp cell using Daphne 7373 silicone oil as the pressure transmitting medium. The pressure was determined, at room temperature, using a manganine resistance gauge located in the pressure cell close to the sample. The values given below take into account the pressure decrease during cooling.

\subsection{Magneto-transport measurements}
\label{sec:1.1}

The analysis of our data is based on the study of the magneto-conductivity and is similar to the one presented in Ref. \onlinecite{Kim93} for multi-carrier semiconductor systems.  The magneto-conductivity is obtained from the measured resistivity tensor by means of $\sigma_{xx}(H)=\rho_{xx}(H)/\left[\rho_{xx}^2(H)+\rho_{xy}^2(H)\right]$.
For a single carrier system, its analytical expression reads\cite{Zakrzewski72,Dziuba74}
\begin{equation}
\sigma_{xx}(H)=\frac{\sigma_{xx}(H=0)}{1+\mu^2 H^2}
\label{sigmaxx2}
\end{equation}
where $\sigma_{xx}(H=0)=e\mu n$, $e$ is the electron charge, $\mu$ the mobility, and $n$ is the carrier density.

\begin{figure}[htb]
	\centering
	\includegraphics[width=0.80\hsize]{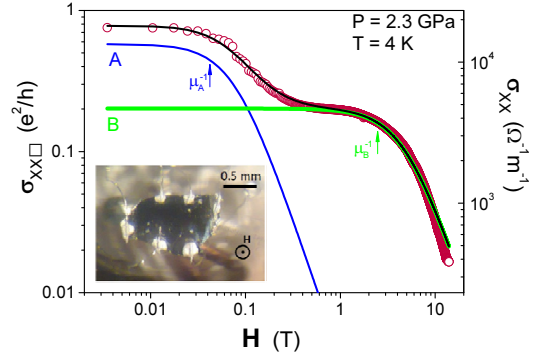}
	\caption{(Color online). Typical magneto-conductivity curve of aI3 (open circles) that can be understood as a two-carrier system (blue line: A-carrier conductance, green line: B-carrier conductance, black line: conductance of a system with both A and B carriers). The mobility of each carrier type is determined as the crossover from a constant conductivity at low fields to the $H^{-2}$ regime at high fields. The left axis shows the square (2D) conductivity of each BEDT-TTF plane while the right axis shows the ``bulk'' (3D) longitudinal conductivity (see text). Inset: Photograph of one sample.}
	\label{magneto}
\end{figure}

Figure \ref{magneto} displays a typical magneto-conductivity curve of aI3 under pressure, where two `plateaus' can be clearly seen. As conductivity in aI3 has a strong 2D character, conductivity is shown both as 3D conductivity ($\sigma_{xx}$) and as 2D conductivity ($\sigma_{xx\square}$ of each BEDT-TTF plane) according to $\sigma_{xx\square}=\sigma_{xx} c$. As conductivity is additive, in a two-carrier system, the contributions of each carrier type A and B can be added,
\begin{equation}
\sigma_{xx}(H)=\frac{\sigma_{xx,A}(H=0)}{1+\mu_A^2 H^2}+\frac{\sigma_{xx,B}(H=0)}{1+\mu_B^2 H^2}
\label{sigmaAB}
\end{equation}

The two ``plateaus'', observed in Fig. \ref{magneto}, indicate the existence of two different carrier types ($\gamma=A$ or $B$)
with significantly different mobilities. From this curve, we can extract the mobilities, $\mu_{\gamma}$, of each carrier type, their zero-field conductivities, $\sigma_{xx,\gamma}(H=0)$, and their carrier densities, $n_{\gamma}$, by $n_{\gamma}=\sigma_{xx,\gamma}(H=0)/e\mu_{\gamma}$.

\begin{figure}[htb]
	\centering
	\includegraphics[width=0.90\hsize]{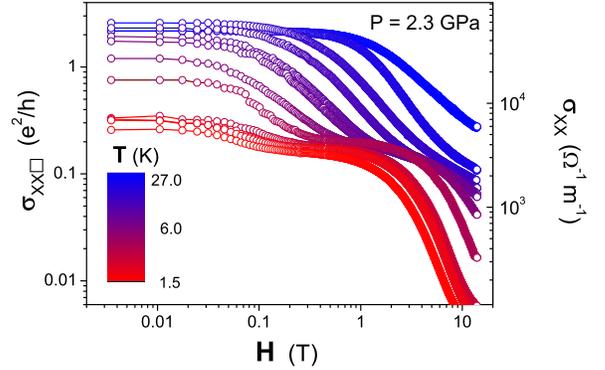}
	\caption{(Color online). Magneto-conductivity curves of aI3 at $P=2.3$ GPa for different temperatures, from bottom to top: $1.5$, $2.2$, $3.0$, $3.9$, $6$, $8$, $9$, $12$, $15$, $20$ and $27$ K. The left axis shows the square (2D) conductivity of each BEDT-TTF plane while the right axis shows the ``bulk'' (3D) longitudinal conductivity.}
	\label{sigmaxxT}
\end{figure}

Figure \ref{sigmaxxT} shows magneto-conductivity curves of aI3 at a fixed pressure for several temperatures. The previous analysis 
has been repeated for each of these magneto-conductivity curves to obtain the densities (Fig. \ref{density})  and mobilities 
(Fig. \ref{mobility}) for each carrier type as a function of temperature and for three different pressures, $P=1.6$, $2.3$ and $3.0$ GPa. 
The strong temperature dependence of the carrier density is a signature that temperature is higher than $T_F$ for both A and B carriers  
even at the lowest measured temperature, $T_F\leq T_{min}=1.5$ K. This low Fermi temperature hints at the absence of charge
inhomogeneities that prevent the approach of the Dirac point in graphene on Si0$_2$ substrates.\cite{Martin08}

\subsection{Temperature dependence of the carrier densities}
\label{sec:1.2}

The carrier density can be calculated from $n_{\gamma}=\int f(E) D_{\gamma}(E) dE$, where $f(E)$ is the Fermi-Dirac distribution and $D_{\gamma}(E)$ is the density of states for massive ($\gamma=M$) and Dirac ($\gamma=D$) carriers:\cite{CastroNeto09}

\begin{figure}[htb]
	\centering
	\includegraphics[width=0.90\hsize]{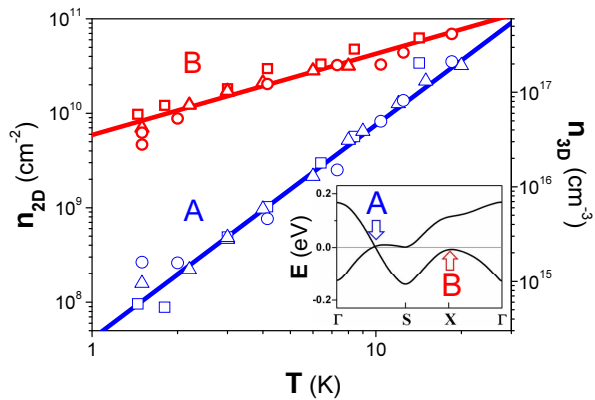}
	\caption{(Color online). A and B carrier densities as a function of temperature (circles: $1.6$ GPa, triangles $2.3$ GPa and squares $3.0$ GPa; blue thin symbols for A carriers and red thick symbols for B carriers). The left axis shows the density for each BEDT-TTF plane ($n_{2D}$) while the right axis shows the bulk density ($n_{3D}$). The lines represent power-law fits of to the A and B carrier densities that yield exponents 0.9 and 2.2, respectively. Inset: band structure calculations at $1.7$ GPa where both Dirac cones (A) and parabolic bands (B) cross the Fermi level (adapted from Ref. \onlinecite{Alemany12}).}
	\label{density}
\end{figure}

\begin{figure}[htb]
	\includegraphics[width=0.70\hsize]{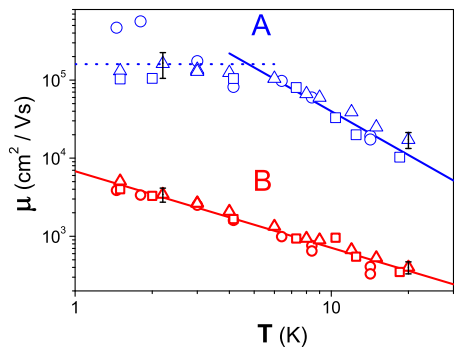}
	\caption{(Color online). Mobilities for A and B carrier types as a function of temperature (circles: $1.6$ GPa, triangles: $2.3$ GPa and squares: $3.0$ GPa; blue thin symbols for A carriers and red thick symbols for B carriers). The continuous lines represent power laws fits for the mobilities dependences with temperature, which gives exponents $-1.9$ (A carriers) and $-1.0$ (B carriers). The low temperature dispersion of A carriers mobility is due to a decrease of the saturating mobility (dotted line) by increasing pressure.}
	\label{mobility}
\end{figure}

\begin{equation}
D_{M}(E)=\frac{g_{v,M} g_s m^*}{2 \pi \hbar^2}
\label{Dmassif}
\end{equation}	

\begin{equation}
D_{D}(E)=\frac{g_{v,D} g_s }{2 \pi \left( \hbar v_F \right)^2} E
\label{DDirac}
\end{equation}	
where $g_{v,\gamma}$ and $g_s$ are valley and spin degeneracies and $m^*$ is the effective mass of massive carriers described by a Schr\"odinger equation. In aI3 under pressure, two Dirac cones but only one massive band exist at the Fermi level.\cite{Alemany12} For large temperatures, $T\gg T_F$, the carrier density depends linearly on temperature  for massive carriers and quadratically for Dirac carriers:

\begin{equation}
n_{M}=\frac{\ln(2) m^*}{ \pi \hbar^2}k_B T
\label{nmassif}
\end{equation}	
\begin{equation}
n_{D}=\frac{2\pi}{ 3} \left(\frac{k_B T}{\hbar v_F}\right)^2 
\label{nDirac}
\end{equation}

Figure \ref{density} represents the measured temperature dependence of the mobilities and reveals a power-law behavior, $n\sim T^{\beta}$. Indeed one obtains an exponent of $\beta\simeq 0.9$ for the low-mobility carriers (B), in good agreement with what [Eq. (\ref{nmassif})] expected for massive carriers, whereas one finds $\beta\simeq 2.2$ for the high-mobility carriers (A), as roughly expected for massless Dirac particles [Eq. (\ref{nDirac})].

\begin{figure}[htb]
	\centering
	\includegraphics[width=0.70\hsize]{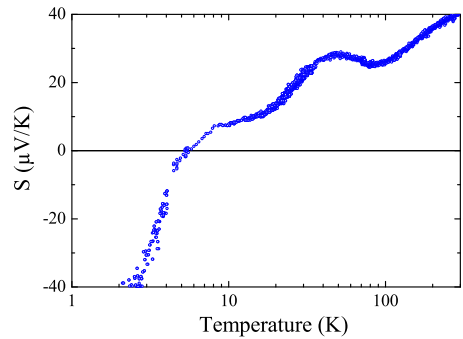}
	\caption{(Color online). Thermopower for a pressure of 1.5 GPa as function of temperature for sample 2. The sign change observed as sweeping temperature confirms the two-carrier picture.}
	\label{thermopower}
\end{figure}

Besides, the nature of the carriers can be extracted from Hall measurements. Furthermore, we have performed thermopower measurements under pressure on a second sample (Figure \ref{thermopower}). These data show a sign change for the Seebeck coefficient (S) around 5K. Thermopower is the voltage per unit of temperature produced by a thermal gradient. The carrier type determines the sign while the density and mobility of the carriers establish the amplitude. Thus, a sign change of the thermopower indicates that the relevant carriers at low temperature have a different charge than those at high temperature, requiring a two-carrier scenario. 

In agreement with Ref. \onlinecite{Tajima12}, A carriers which dominate the low-field conduction are electrons. On the contrary, at large fields the conduction is dominated by holes (B carriers). Notice that our results are consistent with ab-initio calculations of the band structure of aI3 under a pressure of $1.76$ GPa (inset of figure \ref{density})\cite{Alemany12} and do not depend on pressure (within the range $1.6-3.0$ GPa). This supports the idea that massless and massive particles coexist in a broad pressure range. However, since $T>T_F$ in the whole temperature range under study, both Dirac electrons and Dirac holes are excited. Thus there are indeed not two but three carrier types: Dirac holes, Dirac electrons and massive electrons. For $T\gg T_F$, the electron and hole densities are actually identical (semimetal with symmetric band structure): $n_{D,holes}\approx n_{D, electrons}\approx n_{D}/2=n_{A}/2$. The absence of a third `plateau' in the magneto-conductivity data allows us to consider that Dirac electrons and holes have roughly the same mobilities:  $\mu_{D, holes}\approx \mu_{D, electrons}\approx \mu_{D}=\mu_{A}$. Therefore, the results obtained in figure \ref{density} and \ref{mobility} still hold when we consider two types of Dirac carriers (electrons and holes) in addition to the massive holes. This analysis allows us to avoid using Hall effect measurements for the determination of carrier densities. Indeed, Hall effect interpretation becomes challenging as Dirac electron and hole contributions partially compensate, leading to the determination of only an `effective' Dirac carrier density, and they are both mixed with massive carriers contribution. This problem is solved here by analyzing the magneto-conductivity where all carriers contributions are additive.

\subsection{Mobilities}
\label{sec:1.3}

The effective mass of the massive carriers has been extracted from Eq. (\ref {nmassif}). The obtained value is quite small $m^*\approx 0.3\:m_e$ ($m_e$ is the free electron mass). Meanwhile, from Eq. (\ref {nDirac}), $v_F\approx 1.1\times 10^5$ m/s can be extracted, in agreement with previous theoretica\cite{Kobayashi07,Goerbig08} and experimental estimates.\cite{Tajima12} In Fig. \ref{density}, no significant variation of this argument is observed upon sweeping pressure (which should appear as a vertical shift of the $T^2$ line). This indicates that $v_F$ does not change with pressure. In principle, pressure should enhance hopping while reducing the unit cell volume. Thus, an enhancement of the $v_F$ with pressure could be expected according to the approximate expression $v_{F}\simeq ta/\hbar$,
where $t$ is the hopping integral. This expression can be simplified by means of Harrison's law ($t\approx\hbar/m_e a^2$) into $v_{F}\approx \hbar/m_e a$.
As pressure slightly modifies the lattice constant ($1.4 \%/$GPa \cite{Kondo09}), $v_F$ is expected to vary by the same order of magnitude which is smaller than our current experimental uncertainty. This accounts for the apparent absence of pressure effects on the carrier density in the range $1.6 - 3.0$ GPa.

In Fig. \ref{mobility}, the mobility of the Dirac carriers (A) reaches $2\times10^5$ cm$^2$/Vs at low temperatures ($4$ K), a value comparable to 
already published values.\cite{Tajima06} It is quite high compared to typical graphene on \SiO2 values ($10^3$ to $10^4$ cm$^2$/Vs) but similar 
to suspended graphene and graphene on BN mobilities at very low carrier density.\cite{Bolo08, Kim10, kimBN} On the other hand, the mobility 
for massive carriers is $2\times 10^3$ cm$^2$/Vs at $4$ K, which is two orders of magnitude smaller than for Dirac carriers. The temperature 
dependence of the mobility follows power laws for both massive (exponent $-1.0$) and Dirac carriers (exponent $-1.9$). Moreover, the Dirac carrier 
mobility seems to saturate at $T<4K$. A similar saturation has been reported in others Dirac systems.\cite{Chen08}  
Table \ref{tabcomp} summarizes the main parameters of massive and Dirac carriers in aI3, in comparison with graphene on \SiO2 .

\begin{table}
	\centering
\begin{tabular}{|p{2.8cm}|p{1.4cm}|p{1.4cm}|p{2.4cm}|}
\hline
Quantity & Massive carriers in aI3 & Dirac carriers in aI3 & Dirac carriers in graphene/ \SiO2 \cite{Monteverde10}\\ \hline
$T_F$ (K)  & $  <1.5$ & $  <1.5$ & $  \approx 100$ \\ \hline
$n_{2D Minimal}$ (cm$^{-2}$) & $  8 \times 10^9$ & $  2 \times 10^8$ & $  4 \times 10^{11}$ \\ \hline
$v_F$ (m/s) & $  -$ & $  1 \times 10^5$ & $  1 \times 10^6$ \\ \hline
$\mu$(4K) (cm$^2$/Vs) & $  2 \times 10^3$ & $  2 \times 10^5$ & $  10^3-10^4$ \\ \hline
$m^*$ ($m_e$) & $  0.3$ & $  -$ & $  -$ \\ \hline
$\tau$ (fs) & $  200$ & $  300$ & $  75$ \\ \hline
\end{tabular}
\caption{Dirac and massive carriers parameters in aI3 at high pressure, in comparison with graphene electrons. }
	\label{tabcomp}
\end{table}

\section{Theoretical discussion in terms of screened Coulomb or short-range scatterers}
\label{sec:2}

In order to better understand the difference in the mobility, we investigate the ratio $\mu_M/\mu_D=\tau_M m_D/\tau_D m^*$, in terms of the scattering times $\tau_D$ and $\tau_M$ for the massless Dirac and massive carriers, respectively. Furthermore, $m_D=E_{F,D}/v_F^2$ is the density-dependent cyclotron mass of the Dirac carriers, in terms of the Fermi energy $E_{F,D}=k_B T_{F,D}$. The scattering times may be obtained from Fermi's golden rule (for $\gamma=D,M$)
\begin{equation}
\left(\tau_{\gamma}\right)^{-1}=2\pi n_{\rm imp}|V_{\gamma}|^2 D_{\gamma}(E_{F,\gamma}),
\end{equation}
in terms of the impurity density $n_{\rm imp}$, the matrix element $V_{\gamma}$, and the density of states (\ref{DDirac}) for Dirac and (\ref{Dmassif}) for massive carriers. We consider implicitly that both carrier types are affected by the same impurities, and the matrix element is independent of $\gamma$ for short-range impurity scattering. Apart from atomic defects, screened Coulomb-type impurities approximately fulfill this condition, as it may be seen within the Thomas-Fermi (TF) approximation. Indeed, the screening length of the Coulomb interaction is dominated by the Thomas-Fermi wave vector $k_{TF,M}=1/a_0\simeq 10^{10}$ m$^{-1}$ of the massive carriers, for an effective Bohr radius $a_0=\hbar^2/m^* e^2$, whereas the Thomas-Fermi wave vector for massless Dirac carriers $k_{TF,D}=\alpha_D k_{F,D}\sim 10^{8}$ m$^{-1}$, for a density $2\times 10^8$ cm$^{-1}$ and a fine-structure constant $\alpha_D=e^2/\hbar v_F\simeq 20$. The Thomas-Fermi wave vector is thus roughly one order of magnitude larger than the Fermi wave vector of the massive carriers, which is itself much larger than that of the Dirac carriers. The screened Coulomb potential for $\gamma$-type carriers may therefore be approximated by its $q=0$ value, $V_{TF}(q\sim k_{F,\gamma})=2\pi e^2/\epsilon \epsilon_{TF}(q\sim k_{F,\gamma}) q\simeq 2\pi e^2/k_{TF}^B=V_{TF}(q=0)$, which is thus the same for both carrier types, as mentioned above. Here, $\epsilon$ is the permittivity of the dielectric environment and  $\epsilon_{TF}(q)$ is the dielectric function calculated within the Thomas-Fermi approximation.

In view of the above considerations, we thus obtain, for the mobility ratio in the limit $T\rightarrow 0$
\begin{equation}\label{eq:mu}
\frac{\mu_M}{\mu_D}\simeq \frac{2\pi\hbar^2 D_{D}(E_{F,D})}{g_sg_{v,M} m^*}\times \frac{E_{F,D}}{m^* v_F^2},
\end{equation}
which does neither depend on the form of the matrix element nor on the impurity density. One expects a ratio in the $10^{-3}$ range, whereas the measured ratio is $\sim 10^{-2}$ at $T=4$ K. Notice that for $T\gg T_F$, that is in the experimentally relevant regime here, one may replace the energy dependence in the density of states of the massless Dirac carriers by a linear dependence in temperature, $E_{F,D}\rightarrow k_BT$, such that one expects a linear temperature dependence of the mobility ratio (\ref{eq:mu}), in agreement with our experimental findings ($\mu_M/\mu_D \propto T^{0.9}$ for $T>4K$, see Fig. \ref{mobility}).

\section{Conclusions}
\label{sec:3}

To conclude, we present an interpretation of magneto-transport in aI3 that indicates that both massive and Dirac carriers are present even at high pressures. 
Thermopower measurements performed on one of the three studied samples are also in agreement with this two carrier scenario.

 So far in the literature, the conduction in this system has been attributed solely to Dirac carriers.\cite{Tajima06} Moreover, this coexistence holds with little perturbation in the whole range of pressure under study. As Dirac carriers have high mobility, they dominate the conduction at low magnetic field and high temperatures. On the contrary, for high magnetic fields and low temperatures, the massive holes drive the conduction properties. This crossover can be clearly seen from our magneto-conductivity curves and is responsible for their peculiar `plateau' shape. It should also be noted that a proper separation of massive carriers has to be done prior to using any expression that concerns solely Dirac carriers. In order to confirm the picture of coexisting Dirac and massive carriers, complementary studies, such as spectroscopic measurements, are highly desirable but beyond the scope of the present paper.

\begin{acknowledgments}

We acknowledge J.-P. Pouget, H. Bouchiat, G. Montambaux, F. Pi\'echon and J.-N. Fuchs for fruitful discussions.
\end{acknowledgments}


\begin{thebibliography}{9}


\bibitem{Mishima95} T. Mishima, T. Ojiro, K. Kajita, Y. Nishio, and Y. Iye, Synthetic Metals {\bf 70}, 771 (1995). 

\bibitem{Tajima00} N. Tajima, M. Tamura, Y.Nishio, K. Kajita, and Y. Iye, J. Phys. Soc. Jpn. {\bf 69}, 543 (2000).

\bibitem{Tajima06} N. Tajima, S. Sugawara, M. Tamura, Y. Nishio, and K. Kajita, J. Phys. Soc. Jpn. {\bf 75}, 051010 (2006). 

\bibitem{Tajima09} N. Tajima, S. Sugawara, R. Kato, Y. Nishio, and K. Kajita, Phys. Rev. Lett. {\bf 102}, 176403 (2009).

\bibitem{Kino06} H. Kino, and T. Miyazaki, J. Phys. Soc. Jpn. {\bf 75}, 034704 (2006).

\bibitem{Kobayashi07} A. Kobayashi, S. Katayama, Y. Suzumura, and H. Fukuyama, J. Phys. Soc. Jpn. {\bf 76}, 034711 (2007).

\bibitem{Novoselov05} K. S. Novoselov, A. K. Geim, S. V. Morozov, D. Jiang, M. I. Katsnelson, I. V. Gregorieva, S. V. Dubonos, and A. A. Firsov, Nature {\bf 438}, 197 (2005); Y. Zhang, Y.-W. Tan, H. L. Stormer, and P. Kim,
Nature {\bf 438} 201, (2005).

\bibitem{TI}For a review, see M. Z. Hasan and C. L. Kane, Rev. Mod. Phys. {\bf 82}, 3045 (2010); 
X.-L. Qi and S. C. Zhang, Rev. Mod. Phys. {\bf 83}, 1057 (2011).

\bibitem{Goerbig08} M.O. Goerbig, J.-N. Fuchs, G. Montambaux, and F. Pi\'echon, Phys. Rev. B {\bf 78}, 045415 (2008).

\bibitem{CastroNeto09} A.H. Castro Neto, F. Guinea, N.M.R. Peres, K.S. Novoselov, and A.K. Geim, Rev. Mod. Phys {\bf 81}, 109 (2009); M. O. Goerbig
Rev. Mod. Phys {\bf 83}, 1193 (2011); V.  N. Kotov, B. Uchoa, V. M. Pereira, F. Guinea, and A. H. Castro Neto, Rev. Mod. Phys. {\bf 84}, 1067.

\bibitem{Alemany12} P. Alemany, J.-P. Pouget, and E. Canadell, Phys. Rev. B {\bf 85}, 195118 (2012).

\bibitem{Montambaux09} G. Montambaux, F. Pi\'echon, J.-N. Fuchs, and M.O. Goerbig, Phys. Rev. B {\bf 80}, 153412 (2009); Eur. Phys. J. B {\bf 72}, 509 (2009).

\bibitem{Tajima11} M. Sato, K. Miura, S. Endo, S. Sugawara, N. Tajima, K. Murata, Y. Nishio, and K. Kajita, J. Phys. Soc. Jpn. {\bf 80}, 023706 (2011).

\bibitem{Tajima12} N. Tajima, R. Kato, S. Sugawara, Y. Nishio, and K. Kajita, Phys. Rev. B, {\bf 85}, 033401 (2012).

\bibitem{Kim93} J.S. Kim, D.G. Seiler, and W.F. Tseng, J. Appl. Phys {\bf 73}, 8324 (1993).

\bibitem{Zakrzewski72} T. Zakrzewski, and E.Z. Dziuba, Phys. Status Solidi B {\bf 52}, 665 (1972).

\bibitem{Dziuba74} E.Z. Dziuba, Phys. Status Solidi B {\bf 62}, 307 (1974).

\bibitem{Martin08} J. Martin, N. Akerman, G. Ulbricht, T. Lohmann, J. H. Smet, K. Von Klitzing, and A. Jacoby,  Nature Phys. {\bf 4}, 144 (2008).

\bibitem{Kondo09} R. Kondo, S. Kagoshima, N. Tajima, and R. Kato, J. Phys. Soc. Jpn. {\bf 78}, 114714 (2009).

\bibitem{kimBN} C. Dean, A. Young, P. Cadden-Zimansky, L. Wang, H. Ren, K. Watanabe, T. Taniguchi, P. Kim, J. Hone, and K. Shepard, Nature Phys. 7, 693 (2011).

\bibitem{Bolo08} K.I. Bolotin, K.J. Sikes, Z. Jiang, M. Klima, G. Fudenberg, J. Hone, P. Kim, H.L. Stormer, Solid State Communications, {\bf 146}, 351 (2008).

\bibitem{Kim10} C.R. Dean, A.F. Young, I. Meric, C. Lee, L. Wang, S. Sorgenfrei, K. Watanabe, T. Taniguchi, P. Kim, K.L. Shepard, and J. Hone, Nature Nanotechnology {\bf 5}, 722 (2010).

\bibitem{Chen08} J.-H. Chen, C. Jang, S. Xiao, M. Ishigami, and M.S. Fuhrer, Nature Nanotechnology {\bf 3}, 206 (2008).

\bibitem{Monteverde10} M. Monteverde, C. Ojeda-Aristizabal, R. Weil, K. Bennaceur, M. Ferrier, S. Gu\'eron, C. Glattli, H. Bouchiat, J.-N. Fuchs, and D.L. Maslov, Phys. Rev. Lett. {\bf 104}, 126801 (2010).


\end{thebibliography}

\end{document}